\documentclass{article}

\usepackage{amsmath}
\usepackage{amssymb}
\usepackage{graphicx,subfigure,url}
\usepackage{xcolor}
\usepackage{epstopdf}
\usepackage[normalem]{ulem}
\usepackage{mathptmx}

\usepackage{geometry}
\title{3D particle tracking velocimetry using dynamic discrete tomography\footnote{NOTICE: this is the authors version of a work that was accepted for publication in \textbf{Computer Physics Communications}. Changes resulting from the publishing process, such as peer review, editing, corrections, structural formatting, and other quality control mechanisms may not be reflected in this document. Changes may have been made to this work since it was submitted for publication. A definitive version appears in Computer Physics Communications, 187, pp. 130--136, 2015.}}
\author{Andreas Alpers\\
{\small Zentrum Mathematik, Technische Universit\"at M\"unchen} \\ {\small D-85747 Garching bei M\"unchen, Germany}\\
{\small \texttt{alpers@ma.tum.de}}\\
\and
Peter Gritzmann\\
{\small Zentrum Mathematik, Technische Universit\"at M\"unchen} \\ {\small D-85747 Garching bei M\"unchen, Germany}\\
{\small \texttt{gritzman@ma.tum.de}}\\
\and
Dmitry~Moseev\footnote{corresponding author}\\
{\small FOM Institute DIFFER}\\
{\small 3430 BE Nieuwegein, The Netherlands}\\
{\small and}\\
{\small Max-Planck-Institut f\"ur Plasmaphysik}\\ 
{\small D-85748 Garching bei M\"unchen, Germany} \\
{\small \texttt{dmmo@ipp.mpg.de}}\\
\and
Mirko~Salewski\\
{\small Technical University of Denmark, Department of Physics}\\
{\small DK-2800 Kgs. Lyngby, Denmark}\\
{\small \texttt{msal@fysik.dtu.dk}}\\
}

\date{}

\newcommand{\figref}[1]{Fig.~\ref{#1}}
\begin{document}
\maketitle	
\begin{abstract}
Particle tracking velocimetry in 3D is becoming an increasingly important imaging tool in the study of fluid dynamics and combustion as well as plasmas. We introduce a \emph{dynamic discrete tomography} algorithm for reconstructing particle trajectories from projections. The algorithm is efficient for data from two projection directions and exact in the sense that it finds a solution consistent with the experimental data. Non-uniqueness of solutions can be detected and solutions can be tracked individually. 
\end{abstract}

%\begin{keyword}
%PTV \sep particle tracking velocimetry \sep 3D PTV \sep discrete tomography
%\end{keyword}
%\end{frontmatter}

\section{Introduction}
Particle tracking velocimetry (PTV) is a diagnostic technique that plays an important role in studying flows \cite{Adrian1986,Adrian1991,Kieft2002,Suzuki2000,Hadad2013,Ichiyanagi2012,Krug2014,Oliveira2013} including combustion \cite{Reuss1986,Pareja2010,Gurliat2003,Qin2000,Lewis1989,Egolfopoulos2014,Tafnout2013,Wu2013}. It has also been used to study plasma \cite{Hadziavdic2006,Gandy2001,Rosanvallon2008,Williams2011,Williams2007,Oxtoby2012,Krasheninnikov2011,Rudakov2009}. In PTV the motion of particles is followed in a sequence of images to measure their instantaneous velocities. In complex plasmas the particles themselves are the subject of interest \cite{Goertz1989,Shukla2001,Goree2013,Kantor2014,Oxtoby2013} whereas in fluids the particle velocities are nearly the same as the local flow velocities which can hence be studied by PTV. 

PTV is similar to the related particle image velocimetry (PIV) \cite{Adrian1991}. PTV tracks the motion of individual particles whereas PIV tracks the motion of groups of particles statistically. In PTV measurements the concentration of tracer particles is therefore significantly lower than in PIV measurements. In traditional \emph{two-dimensional} PTV or PIV measurements, the flow field is illuminated by a thin laser sheet. Light is scattered from the tracer particles in this laser sheet and imaged on a CCD camera. From a series of images we can then obtain 2D flow velocities in the plane of the laser sheet \cite{Salewski2007,Wang2010}. In stereo PIV measurements the laser sheet is observed with two cameras, and velocities in the plane of the laser sheet can be obtained \cite{Prasad2000}. PIV techniques have been extended to volumetric 3D measurement by scanning planar PIV \cite{Brucker1997}, holographic PIV \cite{Katz2010}, and tomographic PIV \cite{Elsinga2006}. We study the 3D PTV mode of operation in which individual particles are tracked to obtain 3D velocity vectors in a measurement volume \cite{Brucker1997,Pereira2006,Chang1984,maas-93}. The particles either scatter light from a volumetric illumination of the measurement volume or they glow by themselves as often in plasma. 3D PTV \cite{maas-93} is advantageous if the density of particles is intrinsically low or has to be limited.

Current tomographic particle tracking methods are based on the multiplicative algebraic reconstruction technique (MART) \cite{Herman1976} and its variants \cite{Pereira2006,Wieneke2013}. These are methods for reconstructing the distribution of multiple-pixel sized particles  modeled as graylevel images. The graylevel can take any value and is a continuous quantity. The subsequent binarization is usually performed by comparison of the graylevel to a threshold. This procedure is not guaranteed to yield solutions that are consistent with the data.  In contrast, our algorithm returns binary solutions that are consistent with the data as this is explicitly included as a constraint in the imaging model. Information from previously reconstructed frames is incorporated in the reconstruction procedure that is formulated as a \emph{discrete optimization problem}. To our knowledge, discrete optimization methods have not previously been applied in PTV.

Existing PTV algorithms (such as \cite{kitzhofer-10,jon-12,Guezennec1994}) rely on the following assumptions: (a) Applied reconstruction routines are computationally efficient; (b) The reconstructions are stable, i.e., reconstruction errors are small whenever measurement errors are small; and (c) The reconstructions are uniquely determined by the data. The algorithms are therefore generally not able to deal with ambiguities in the reconstruction and typically require heuristic knowledge for particle tracking. Many PTV algorithms, including the one presented in this paper, utilize information from previously reconstructed frames \cite{Novara2011,Schanz2013,Xu2008,Willneff2003}.

Here we discuss efficiency, stability, and uniqueness of the trajectory reconstructions in 3D PTV by relating them to results from the mathematical field of \emph{discrete tomography}, which has originally been developed for reconstructing crystalline objects from high-resolution transmission electron microscopy (HRTEM) data \cite{sksbko-93}; see also \cite{kubaherman1,gritzmann-devries-2001,kubaherman2}. Discrete tomography is preferred over conventional computer tomography (CT) in such tasks, because CT algorithms are, firstly, not well-suitable for reconstructing distributions of pixel-sized objects and, secondly, well-known to generate severe artefacts in cases where projection data is available only from a few directions. 

We introduce a \emph{dynamic} discrete tomography algorithm for 3D PTV, which can efficiently reconstruct trajectories of pixel-size objects from projection data acquired from two directions. The projections are assumed to be acquired along lines, i.e., two 1D detectors are required for particles that are confined to a plane (which could be also called 2D PTV) whereas two 2D detectors are required for particle tracking in 3D. Performing reconstructions from only a few projections can be important in experimental set-ups with limited optical access. For example, in machines for studying high-temperature plasmas the available space for diagnostics is usually very limited and possibilities of reducing the amount of in-vessel equipment are beneficial \cite{Krasheninnikov2011,Rudakov2009,Goodall1982,Saito2007}.

Another potential application of the 3D PTV algorithm is a recent experiment on a gliding arc \cite{Sun:13, Zhu2014}. A gliding arc is a thin string-like plasma column that is suspended between two electrodes while it is convected in a turbulent free jet \cite{Fridman1998,Pellerin2000,Richard1996}. The gliding arc can be used in surface treatment (adhesion) \cite{Kusano2013}, bacterial inactivation \cite{Du2012}, and many other applications. It has been found by PTV \cite{Richard1996} and by measurements with a Pitot tube \cite{Pellerin2000} that the jet flow is about 10-20\% faster than the plasma column. Spatial resolution of the slip velocity (i.e., the velocity of the jet flow measured relatively to the velocity of the plasma column) is not available in the literature as the seeding density of particles was too low. In the gliding arc experiment the density of seed particles should not be too high as the plasma column might otherwise be disturbed. Further, to study the gliding arc, images at a frame rate of 420 kHz and a resolution of $64 \times 128$ pixels have been used \cite{Sun:13}. The pixel resolution was this low for the benefit of the high frame rate. 

We introduce our imaging model in Sect.~\ref{sect:model}, present our dynamic discrete tomography algorithm for 3D PTV in Sect.~\ref{sect:algorithm}, and discuss stability and uniqueness of the solutions in Sect.~\ref{sect:uniqueness}. Performance of the algorithm is demonstrated in Sect.~\ref{sect:testtracking}, followed by the conclusions in Sect.~\ref{sect:conclusions}.

\section{Imaging Model}\label{sect:model}
We assume that one-dimensional projections of the particles are acquired from at least two projection directions (i.e., projections, either in 2D or 3D, are acquired along lines from at least two directions). The number of projection directions is henceforth denoted by $m$. In 3D PTV applications, a projection can be understood as a mapping from 3D space to 2D space, i.e. from real space to a photo image. Similarly, an analogous mapping from 2D space to 1D space can be considered if the sample is confined to a plane. The projection can be represented by binary-valued functions where $1$ represents detection of a particle and $0$ represents non-detection. We remark that this differs from PIV and computerized tomography cases in which intensities are measured that can take any value and that are therefore continuous quantities. In PTV applications, however, it is challenging to relate the detected brightness level to the number of particles lying on the corresponding projecting lines. The binary-valued data, on the other hand, are readily available. 

A parallel beam geometry as indicated in some of the figures is not essential in our case. For $m$ projection directions, $m$ projecting lines pass through every particle. The intersections of these projecting lines for every projection direction are called \emph{candidate points}. The set of candidate points is the so-called \emph{(candidate) grid}; it contains the set of all particle positions and typically many additional points that are all other intersections of these projecting lines. We assume throughout the paper that we have $n$ particles. Hence we have at most $n$ projecting lines for each projection direction, and thus the number of grid points in the corresponding 2D grid does not exceed $n^2$. However, the grid can differ in different time steps since multiple particles might be lying on a projecting line. Also note that the grid can be computed efficiently from the data by computing the intersection points of the corresponding projecting lines. 

\begin{figure}[htb]
\begin{center}
\subfigure[]{\includegraphics[width = 60mm]{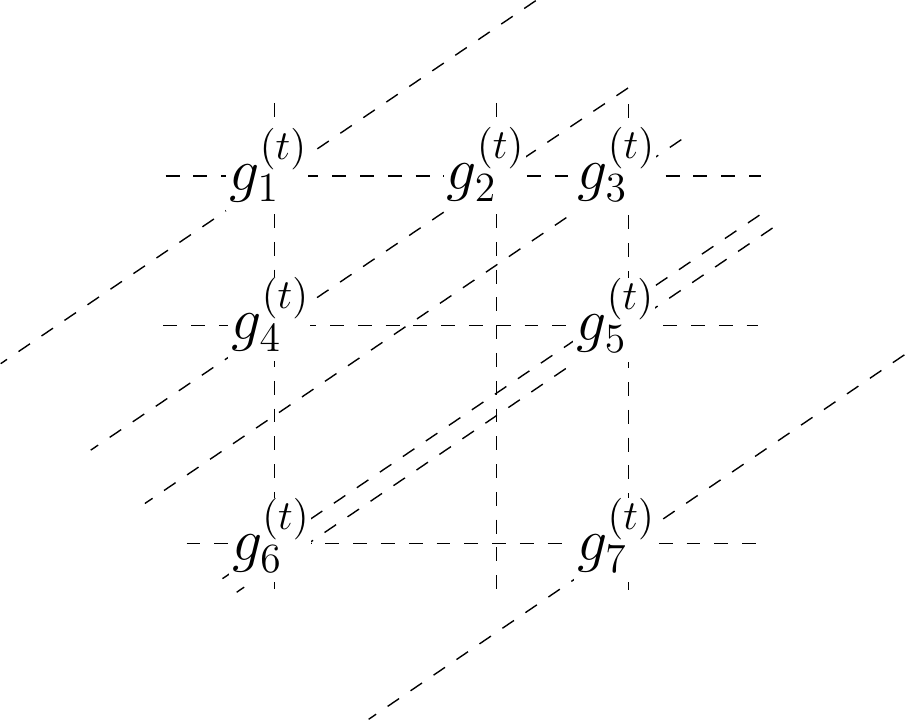}}\hspace*{3ex}
\subfigure[]{\includegraphics[width = 60mm]{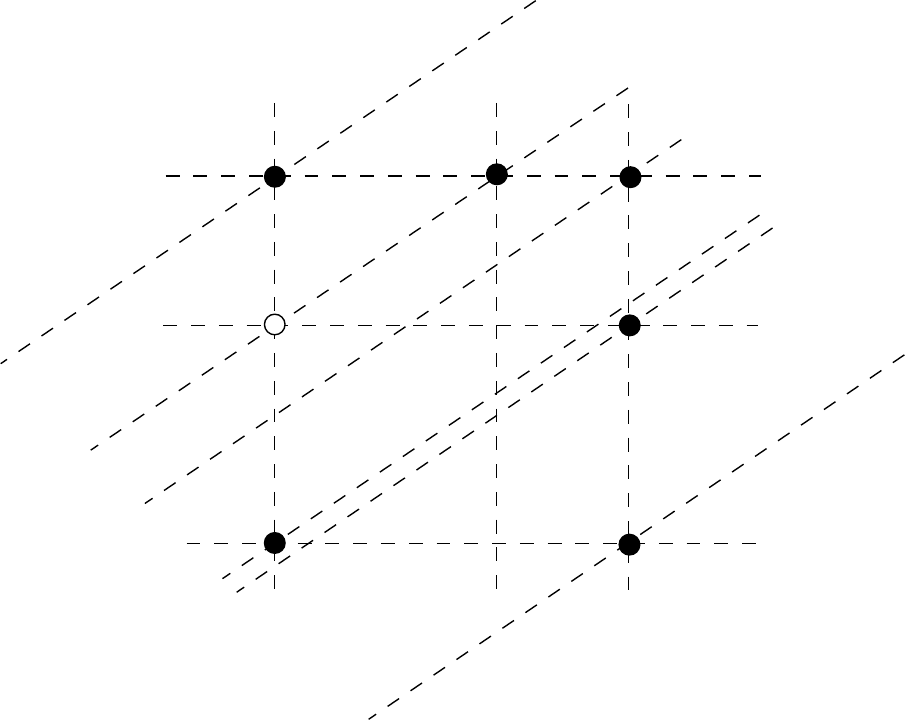}}
\caption{A two-dimensional example for three projection directions (signal-recording lines indicated by dashed lines). (a) The grid $G^{(t)}$, and (b) a possible solution $\vec {\underline{x}}^{(t)}$ representing a set of particles that are consistent with the projections (black and white dots corresponding to $\xi^{(t)}_i=1$ and $\xi_4^{(t)}=0$, respectively).}
\label{fig:grid}
\end{center}
\end{figure}

We consider now the reconstruction problem at time $t$. To each point $g_{i}^{(t)}$ of the candidate grid $G^{(t)}$ containing $l(t)$ points we associate a variable $\xi_{i}^{(t)}$. Presence or absence of a particle at $g_i^{(t)}$ is indicated by the value $\xi_{i}^{(t)}=1$ and $\xi_{i}^{(t)}=0$, respectively; see also Fig.~\ref{fig:grid}.
The requirement that any solution $\vec {\underline{x}}^{(t)}:=(\underline{\xi}_{1}^{(t)},\dots,\underline{\xi}_{l(t)}^{(t)})^T \in \{0,1\}^{l(t)}$ obtained by a reconstruction algorithm should be consistent with the projection data can be described by a \emph{0-1-system of linear inequalities}: 
\begin{equation}\label{eq:model:eq1}
A^{(t)}\vec x^{(t)}\geq \vec b^{(t)}, \quad \vec x^{(t)}\in\{0,1\}^{l(t)},
\end{equation} 
where  $\vec b^{(t)}:=(1,\dots,1)^T\in \{1\}^{k(t)}$ represents the data; $k(t)$ denotes the total number of measurements,  and $A^{(t)}\in\{0,1\}^{k(t)\times l(t)}$ collects the individual variables'  contributions to the signal as specified by the acquisition geometry (for the top horizontal line in Fig.~\ref{fig:grid}, for instance, we would have $\xi_1^{(t)}+\xi_2^{(t)}+\xi_3^{(t)}\geq 1$). Note that we distinguish in our notation between variables $\vec x^{(t)}$ and particular solutions $\vec{\underline{x}}^{(t)}$. 

If no projecting line contains two particles, we can reformulate~\eqref{eq:model:eq1} as 
\begin{equation}\label{eq:model:eq2}
A^{(t)}\vec x^{(t)}=\vec b^{(t)}, \quad \vec x^{(t)}\in\{0,1\}^{l(t)}.
\end{equation} 
Integer vectors $\vec b^{(t)}$ with entries greater than $1$ are also possible in this framework if the brightness level can be related to the number of particles on projecting lines. We restrict our exposition, however, to the model presented in~\eqref{eq:model:eq1}. 

For the tracking problem, we need to solve~\eqref{eq:model:eq1} for subsequent time steps and need to be able to match the particles from $\vec {\underline{x}}^{(t-1)}$ to the particles from the $\vec {\underline{x}}^{(t)}$ solution. 

\section{A dynamic discrete tomography algorithm for PTV}\label{sect:algorithm}
We introduce a discrete tomography method, which is exact in the sense that it is guaranteed to yield a solution that matches the data. This method is dynamic since the algorithm uses the solution from the previous time step. Non-uniqueness of solutions can be detected via this method, and all solutions can be tracked individually if their number remains small. The method is hence capable of dealing with data insufficiency leading to \emph{ghost} particles, which constitute spurious solutions that are not the real solution but are consistent with the recorded data. Elimination of ghost particles can often be performed at a later stage based on physical arguments. Knowledge of the initial particle positions is not required, but any knowledge of a particle position during the tracking (possibly by relying on additional measurements) can potentially reduce the number of alternative reconstructions to be followed. 

Let $\vec w^{(t)}=(\omega_{1}^{(t-1,t)},\dots, \omega_{l(t)}^{(t-1,t)})^T \in \mathbb{R}_+^{l(t)}$
 denote a vector specifying weights associated to each grid point (possible choices are discussed below). We introduce the following \emph{discrete optimization problem} for the tracking step from $t-1\to t$:
\begin{eqnarray}\label{eq:optproblem}
\textnormal{minimize } & \vec w^{(t)} \bullet \vec x^{(t)},\nonumber\\
\textnormal{subject to } &A^{(t)}\vec x^{(t)}\geq\vec b^{(t)},\\
&\vec x^{(t)}\in\{0,1\}^{l(t)},\nonumber
\end{eqnarray}
where $\vec w^{(t)} \bullet \vec x^{(t)}$ denotes the scalar product between $\vec w^{(t)}$ and $\vec x^{(t)}$. This is a \emph{rolling horizon} approach; a \emph{full horizon} approach is also possible and can potentially reduce further ambiguities.

One possible choice for $\vec w^{(t)}$ is
\begin{equation}\label{eq:m1:1}
\omega_{i}^{(t-1,t)}:=\min_{j:\xi^{(t-1)}_{j}=1}\{\textnormal{dist}(g_{i}^{(t)},g_{j}^{(t-1)})\},
\end{equation}
with $\textnormal{dist}(g_{i}^{(t)},g_{j}^{(t-1)})$ denoting the distance (possibly but not necessarily Euclidean) between the two grid points $g_{i}^{(t)}$ and $g_{j}^{(t-1)}$. Note that $\xi^{(t-1)}_{j}=1$ indicates that a particle is located at grid point $g_{j}^{(t-1)}$. The algorithm thus prefers to fill candidate points that are (in some sense depending on $\textnormal{dist}$) close to particles from the previous time step. If the initial distribution of particles is unknown, we can set $\vec w^{(0)}:=\vec{0}$ thereby giving no preference to any position. Alternative solutions can be found as described later. 

The Euclidean distance function is a suitable choice for slowly moving particles, i.e.,  
\[
\textnormal{dist}(g_{i}^{(t)},g_{j}^{(t-1)}):=||g_{i}^{(t)}-g_{j}^{(t-1)}||_2,
\] where the concept of slow motion has to be understood relative to the frame rate. Using modern high-speed cameras with frame rates of MHz, such a choice of the weighting function can be relevant to a large number of fluid dynamics experiments.

The momentum information can also be incorporated into the weights. If the particles, for instance, are known to move with a certain velocity, then a possible choice would be
\begin{equation}\label{eq:distfunction}
\textnormal{dist}(g_{i}^{(t)},g_{j}^{(t-1)}):=\left\{\begin{array}{lll}c_1,&\textnormal{ for}&r_1>||g_{i}^{(t)}-g_{j}^{(t-1)}||_2,\\c_2,&\textnormal{ for}& r_1\leq ||g_{i}^{(t)}-g_{j}^{(t-1)}||_2\leq r_2,\\c_3,&\textnormal{ for}&r_2<||g_{i}^{(t)}-g_{j}^{(t-1)}||_2,\end{array}\right.
\end{equation}
where $r_1$, $r_2$, $c_1$, $c_2$, $c_3$ are prescribed non-negative numbers with $c_2<\min\{c_1,c_3\}$. A particle at $g_{j}^{(t-1)}$ thus most likely moves a distance between $r_1$ and $r_2$; no displacement direction is preferred in this example. The distance $r_1$ can also be set to zero, which implies that the particle moves most likely a distance smaller than $r_2$ in any direction. This particular case could be a model for the random walk of a particle in a turbulent flow. Knowledge about displacement direction ranges can also be incorporated.

It should be noted, however, that in case of multiple solutions to~\eqref{eq:model:eq1} it may happen that optimal solutions to~\eqref{eq:m1:1} contain ghost particles. Nevertheless, non-uniqueness can be detected in this framework. The solution $\vec {\underline{x}}^{(t)}$ is non-unique if and only if the minimal value of $\vec {\underline{x}}^{(t)}\bullet \vec y^{(t)}$ of the following optimization problem is smaller than the number of particles $n$:
\begin{eqnarray}
\textnormal{minimize } & \vec {\underline{x}}^{(t)}\bullet \vec y^{(t)},\nonumber\\
\textnormal{subject to } &A^{(t)}\vec y^{(t)}\geq\vec b^{(t)},\label{eq:nonuniqueness}\\
&\vec y^{(t)} \in\{0,1\}^{l(t)},\nonumber
\end{eqnarray}
This is the same type of optimization problem as in~\eqref{eq:optproblem}, now with $\vec y^{(t)}$ representing the variables leading to another possible reconstruction.

Moreover, we can check whether there exist solutions that avoid prescribed sets $G'\subseteq G^{(t)}$ of candidate positions. The problem
\begin{eqnarray*}
\textnormal{minimize } & \vec w\bullet \vec y^{(t)},\nonumber\\
\textnormal{subject to } &A^{(t)}\vec y^{(t)}\geq\vec b^{(t)},\\
&\vec y^{(t)}\in\{0,1\}^{l(t)}
\end{eqnarray*}
with $\vec w=\left(\omega_{1},\dots,\omega_{l(t)}\right)$ and
\[
\omega_{i}=\left\{\begin{array}{ll}1, &\textnormal{ for }g_{i}^{(t)}\in G',\\ 0,&\textnormal{otherwise},\end{array}\right.
\]
has objective function value zero if and only if there is a solution avoiding $G'$. Note that combinations with~\eqref{eq:m1:1} are also possible.

The discrete optimization problem in~\eqref{eq:optproblem} can be solved efficiently (i.e., in polynomial time rather than in exponential time) for data taken from two projection directions \cite[Chapter~2]{kubaherman1}. By~\eqref{eq:nonuniqueness} it is in this case thus also possible to determine in polynomial time whether the solution is unique. The ambiguity causing structures (so-called \emph{switching components}) are well understood \cite{langfeld11}, \cite[Chapter~3]{kubaherman1}). In fact, if the number of solutions for each frame is bounded by some constant $C$, then it is possible to determine all  solutions in the two projection direction case for a given frame in $O(Cn^4)$ time \cite{alpersgritzmann13} (the running times of naive approaches are exponential in $n$). 

A worst-case performance of $O(n^3\log(n\max_{i}\{\omega_{i}^{(t-1,t)}\}))$ for solving~\eqref{eq:optproblem} with $m=2$ is guaranteed by the \emph{cost scaling algorithm} \cite{networkflow}. Also by the \emph{simplex method} \cite{Dantzig} it is possible to find so-called \emph{vertex solutions} $\vec {\underline{x}}^{(t)}$ of the \emph{linear program}
\begin{eqnarray*}\label{eq:optproblem2}
\textnormal{minimize } & \vec w^{(t)} \bullet \vec x^{(t)},\nonumber\\
\textnormal{subject to } &A^{(t)}\vec x^{(t)}\geq\vec b^{(t)},\\
&0 \leq \xi_i^{(t)}\leq 1,\quad i=1,\dots,l(t).\nonumber
\end{eqnarray*} For two projection directions, due to the structure of $A^{(t)}$, it is guaranteed that a vertex solution $\vec {\underline{x}}^{(t)}$ is binary and thus solves~\eqref{eq:optproblem}. Popular and well-known linear (and integer linear) programming solvers include CPLEX and Xpress. See \cite{optsoftware} for a list of both commercial and non-commercial solvers.

The situation for $m\geq3$ is different. It can be shown that solving the discrete optimization problem~\eqref{eq:optproblem} for $m\geq 3$ projection directions is at least as hard as finding solutions to the notoriously hard  \emph{traveling salesman problem} \cite{ggp-99}. The problems are said to be NP-hard (for a general overview see \cite{gareyjohnson79}). Note that NP-hardness is a problem-related property; any PTV method that attempts to reconstruct particles from $m \geq 3$ projection directions is affected by this. Small problem instances, however, can in practice be solved via integer linear programming. Approximation algorithms and guarantees are discussed in \cite{gpvw-98}; for the detailed mathematical analysis of PTV problems in the discrete tomography context see \cite{alpersgritzmann13}.

\section{Uniqueness and Stability}\label{sect:uniqueness}
\begin{figure}[htb]
\begin{center}
\subfigure[]{\includegraphics[width = 17mm]{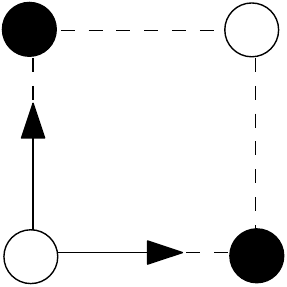}}\hspace*{12ex}
\subfigure[]{\includegraphics[width = 30mm]{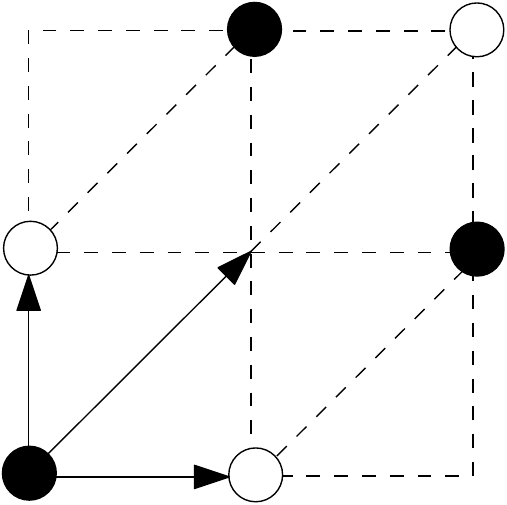}}
\caption{Non-unique reconstructions from (a) two and (b) three projection directions.  Black and white circles denote two possible solutions with identical projection data; projection directions are indicated by arrows.  Similar examples exist for any prescribed set of projection directions.}
\label{fig:switchings}
\end{center}
\end{figure}

It cannot be expected in general that the projections determine solutions uniquely. In fact, for any finite number of projections there is the possibility of non-uniqueness. This is indicated in Fig.~\ref{fig:switchings} for $m\in\{2,3\}$ and can be demonstrated for any $m\geq 2$; see \cite[Chapter~3 and~4]{kubaherman1}. It should be noted that these examples with few projection directions fit into small bounding boxes, i.e., non-uniqueness can appear in many PTV applications even if the sample volume is rather confined.   

\begin{figure}[htb]
\begin{center}
\includegraphics[width = 0.9\textwidth]{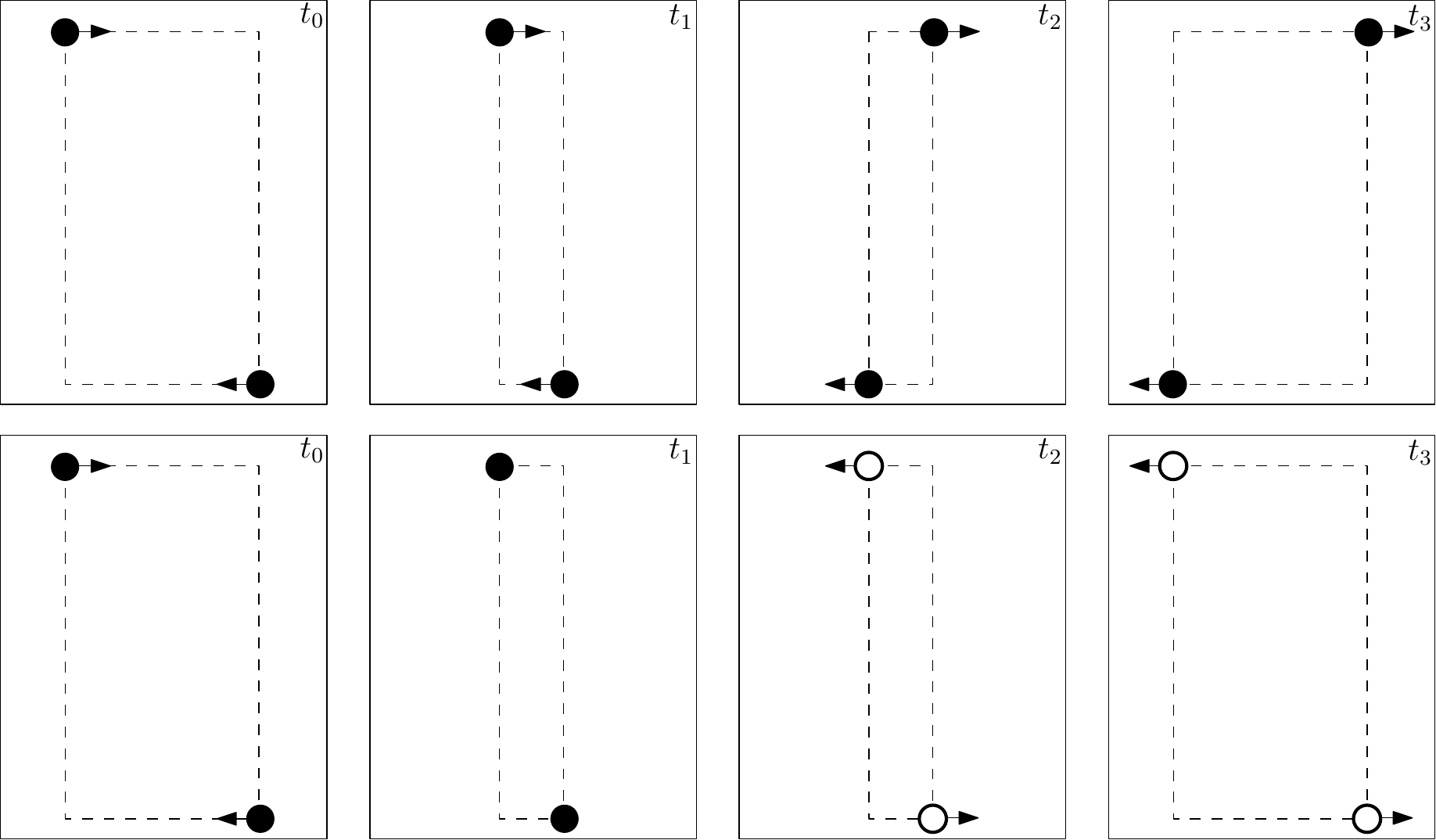}
\caption{An example illustrating that PTV may track particle paths that do not exist but which are consistent with the data.}
\label{fig:switch}
\end{center}
\end{figure}

Prior information in PTV is often of the form that particle movement between frames is assumed to be restricted. Our example in Fig.~\ref{fig:switch} shows that non-uniqueness can occur even in this case. The top row of Fig.~\ref{fig:switch} shows two particles moving into opposite directions; projections are taken along horizontal and vertical directions; the vertical distance between the two particles can be arbitrarily large. The second row shows a different particle movement (i.e., a second solution) that satisfies the same projection data and the same maximum particle displacement as in the top row. (More involved and physically realistic examples are given later.) This also shows that the possibility of non-uniqueness at only two consecutive frames at times $t_1$ and $t_2$ (even if initial conditions at $t_0$ are known), can lead to the tracking of ghost particles, which, over time, may lead to the reconstruction of completely different particle tracks. Note that no second solution in the respective row at $t_3$ is possible by our assumption of restricted particle movement between frames. The example extends to cases with $m >2$ as switching components exist also in these cases (see, e.g., \cite[Chapter~4]{kubaherman1}).

The upshot of our discussion is that it can be desirable to have an algorithm at hand that can generate all solutions, such as presented in Sect.~\ref{sect:algorithm}. Not only physical assumptions can be verified or falsified in this way, the algorithm can also form the basis for a subsequent solution elimination step based on more sophisticated types of prior information (as, for instance, employed in \cite{malik-93}). Individual tracking of non-unique solutions may be implemented within a parallel computing architecture. 

We conclude this section by summarizing several results from the field of discrete tomography that can be applied to PTV reconstructions. The results are sharp in the sense that there are non-unique solutions if the premisses of the results are weakened.

A classic result due to R{\'e}nyi \cite{renyi-52} and its generalization due to Heppes \cite{heppes-56}, for instance, guarantees uniqueness from $m=n+1$ projections if the number of particles is at most $n$. 

A result independent of $n$ but depending on the grid size, guarantees uniqueness if some of the angles between the projection directions are small~\cite{alpersgritzmann13}. Unique reconstructions of particles in a $512\times512$ grid spanned by two orthogonal projection directions, for instance, are guaranteed if the angle between a third and the first projection direction is at most $0.112$ degrees. 

Particle configurations with geometric structure, in particular convex sets of particles, can be uniquely reconstructed from a small number of projections \cite{gardner-gritzmann-97}. Other results can be found in \cite{daurat}; for further references, see \cite[Chapter~4]{kubaherman1}. 

Reconstruction tasks in discrete tomography can potentially be highly unstable in the following sense.  For any set of $m\geq 3$ prescribed projection directions we can find two sets of arbitrarily large cardinality with the following properties: (i)~the sets are uniquely determined by their projections, (ii)~the projections differ on only $2(m-1)$ projecting lines, and (iii)~the sets are disjoint \cite{agt-01}. An error on four projecting lines can thus already lead to disjoint reconstructions in the case of $m=3$ projection directions. Stability results are available \cite{alpers-gritzmann-06} for projection differences smaller than $2(m-1)$ and in much weaker form for uniquely determined sets from $m=2$ projection directions \cite{alpers-brunetti-07,dahlen09}. Stability thus depends strongly on the particular reconstruction scenario; generally it is advisable to incorporate prior knowledge into the reconstruction process (see discussion in Sect.~\ref{sect:algorithm}). 

\section{Test tracking}\label{sect:testtracking}
We test the algorithm presented in Sect.~\ref{sect:algorithm}, and especially its capability to detect and track non-unique solutions, on two test cases based on synthetic data. Test Case~I shows the ability to track non-unique solutions. Test Case~II demonstrates the ability of dealing with 3D motion of particles and shows that the change of dimensionality does not influence the efficiency of the algorithm. This is only meant as an initial proof-of-concept. Extended practical tests based on physical measurements are currently in preparation. 

\subsection{Test Case I}
The 2D data for Test Case~I was generated by Vedenyov's 2D gas simulation MATLAB package \cite{gassimulator}. Six particles (with integer coordinates) were placed into a $70\times70$ pixel grid representing the sample area. Motion of the particles, which resembles diffusion in two dimensions, was followed over $50$ frames. The average particle displacement between consecutive frames was $1$ pixel. 

For solving~\eqref{eq:optproblem} we chose $\vec w^{(t)}$ as in~\eqref{eq:m1:1} and the distance function as in~\eqref{eq:distfunction} with $r_1=\sqrt{2}$, $r_2=2\sqrt{2}$, $c_1=1$, $c_2=2$, and $c_3=60$. Preference is thus given to inter-frame motion within a radius of $r_1$. Projections were taken from two directions, along lines parallel to the $x$-axis and $y$-axis, respectively.

\begin{figure}[htb]
\begin{center}
\subfigure[]{ \scalebox{0.6}{\includegraphics{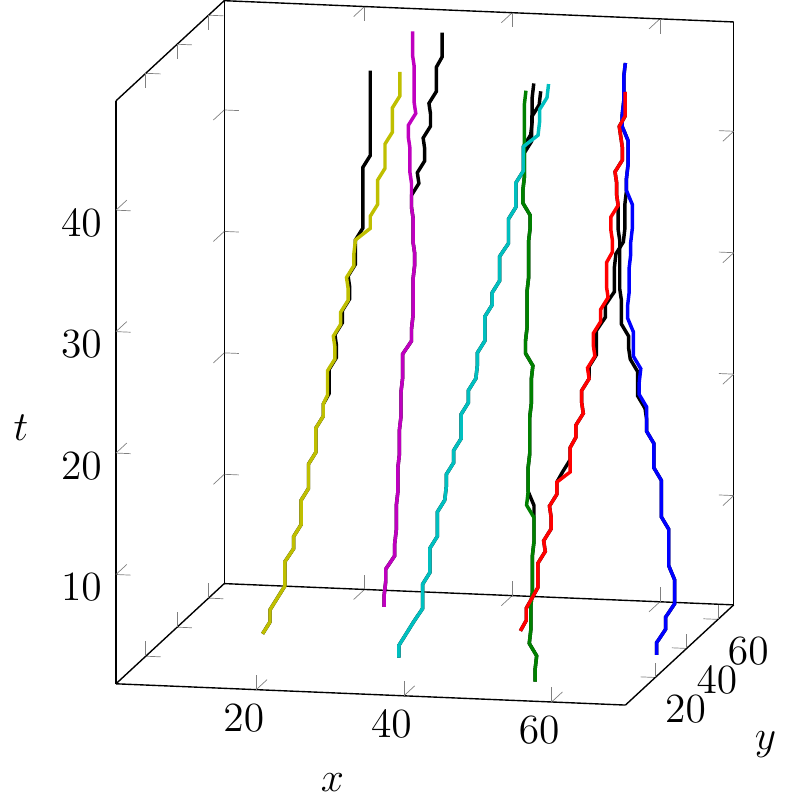}}}\hspace*{0ex}
%\subfigure[]{ \scalebox{0.7}{\input{tracked1.tikz}}}\hspace*{0ex}
\subfigure[]{ \scalebox{0.4}{\includegraphics{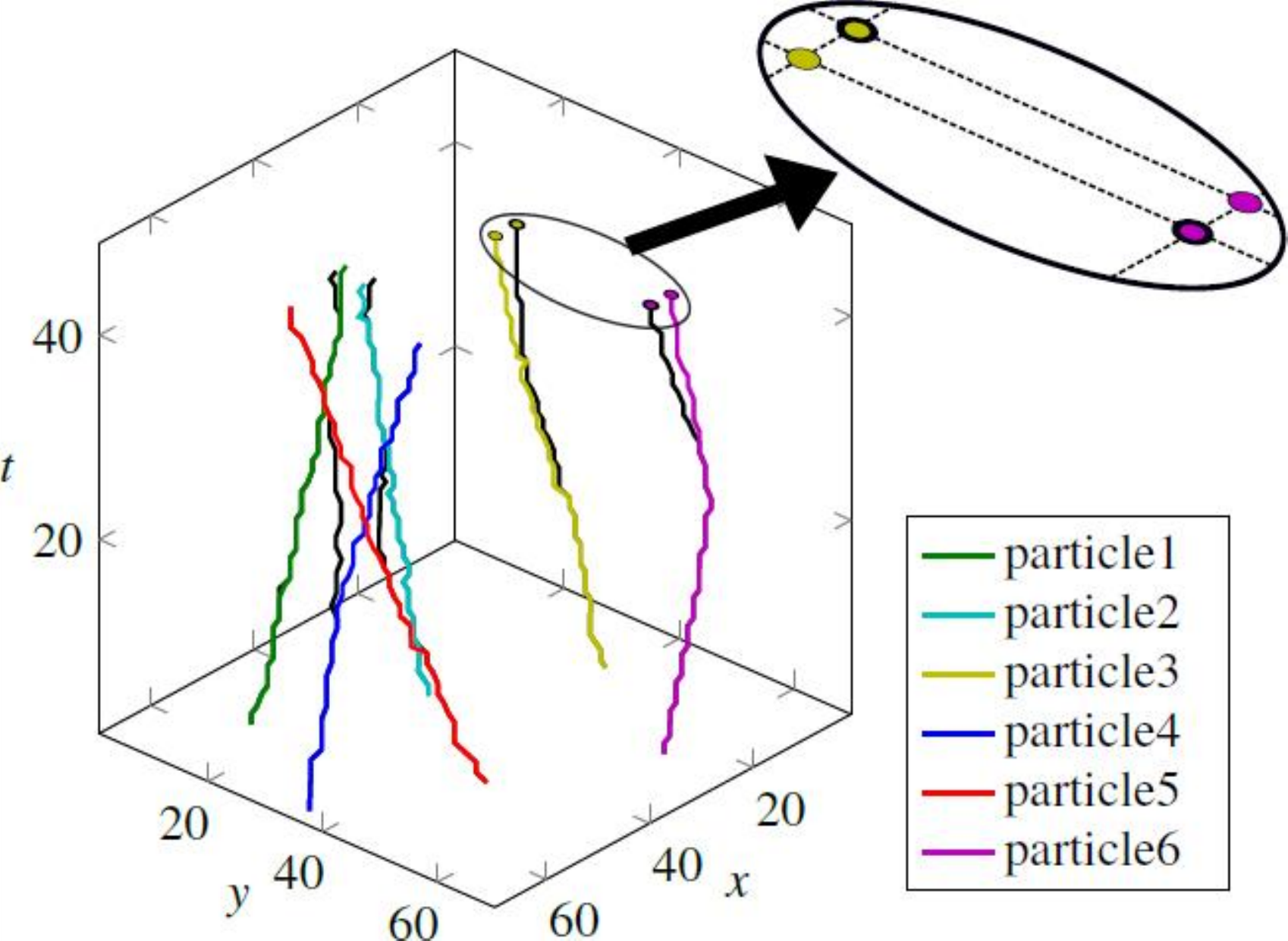}}}
\end{center}
\caption{Tracking of $6$ particles in 2D over $50$ frames; results are shown from two viewing angles. Original particle paths are colored according to the color scheme shown in~(b). Reconstructions deviating from the original are depicted in black color. Particle movement was in two dimensions. The zoomed part shows the deviations of the tracks for Particle~3 and 6 as explained in connection with Fig.~\ref{fig:switch}.}\label{fig:6particles}
\end{figure}

\figref{fig:6particles} shows the tracking results from two different viewing angles with the $t$-axis extending into the third dimension. The real particle paths are colored, and the algorithm finds the real particle paths from the projections. It also finds alternative particle paths (ghost particles), and these deviating particle tracks are colored black and show the non-uniqueness of the problem. (In other situations, the ghost particles may be the real particles, and vice versa.) The first deviations appear in Frame~16 for Particle~1 and~5. Deviations in the reconstruction of Particle~3 and~6 can also be seen. As explained in Sect.~\ref{sect:uniqueness}, these deviations occur since two particles cross a common projecting line; see ~\figref{fig:switch} and the zoomed part of Fig.~\ref{fig:6particles}. The average deviation per particle remains here below $4$ pixels. 

\subsection{Test Case II}
For 3D tracking with heterogeneous particle densities, we observed $500$ particles in a $640\times 640 \times 480$ measurement volume and calculated the photo images that would be recorded by two perpendicularly viewing cameras with standard VGA resolution of $640\times480$; the imaging planes were the $(x,z)$ and $(y,z)$-plane, respectively. We simulated a velocity field with a radial component orthogonal to a line $\ell$ and proportional to $1/r$, where $r$ is the distance to the line $\ell$ that acts as a repellor. The line $\ell$ is obtained by rotating the line parallel to the $z$-axis that passes through the center of the volume by $45$ degrees around the $x$-axis. The velocity field also had an axial component along $\ell$ such that the velocity magnitude is constant; the $z=240$ plane rotated by $45$ degrees around the $x$-axis serves as an attractor. The local particle density in the volume is heterogenous; a higher concentration of particles is found in the vincinity of $\ell$. The average particle displacement between frames was $7.2$ pixels. Whenever new particles enter the measurement volume, we assume that their positions are known for that particular (but no subsequent) frame. This is a rather mild restriction reflecting the fact that the particle density outside the measurement volume is small (the solution in Frame~$1$, for instance, is unique).
 
For solving~\eqref{eq:optproblem} we chose $\vec w^{(t)}$ as in~\eqref{eq:m1:1} and the distance function as in~\eqref{eq:distfunction} with $r_1=6$, $r_2=8$, $c_1=1$, $c_2=1$, and $c_3=9$. Preference is thus given to inter-frame motion within a radius of $r_2$.

We determined the number of different solutions for the reconstruction problem $A^{(t)}\vec x^{(t)}\geq \vec b^{(t)}$, $\vec x^{(t)}\in\{0,1\}^{l(t)}$ at each time step (see Sect.~\ref{sect:algorithm}); the numbers range from $1$ in the first frame to $4.1\cdot 10^{62}$ in the last frame.

\begin{figure}[htb]
\begin{center}
\includegraphics[width = 0.9\textwidth]{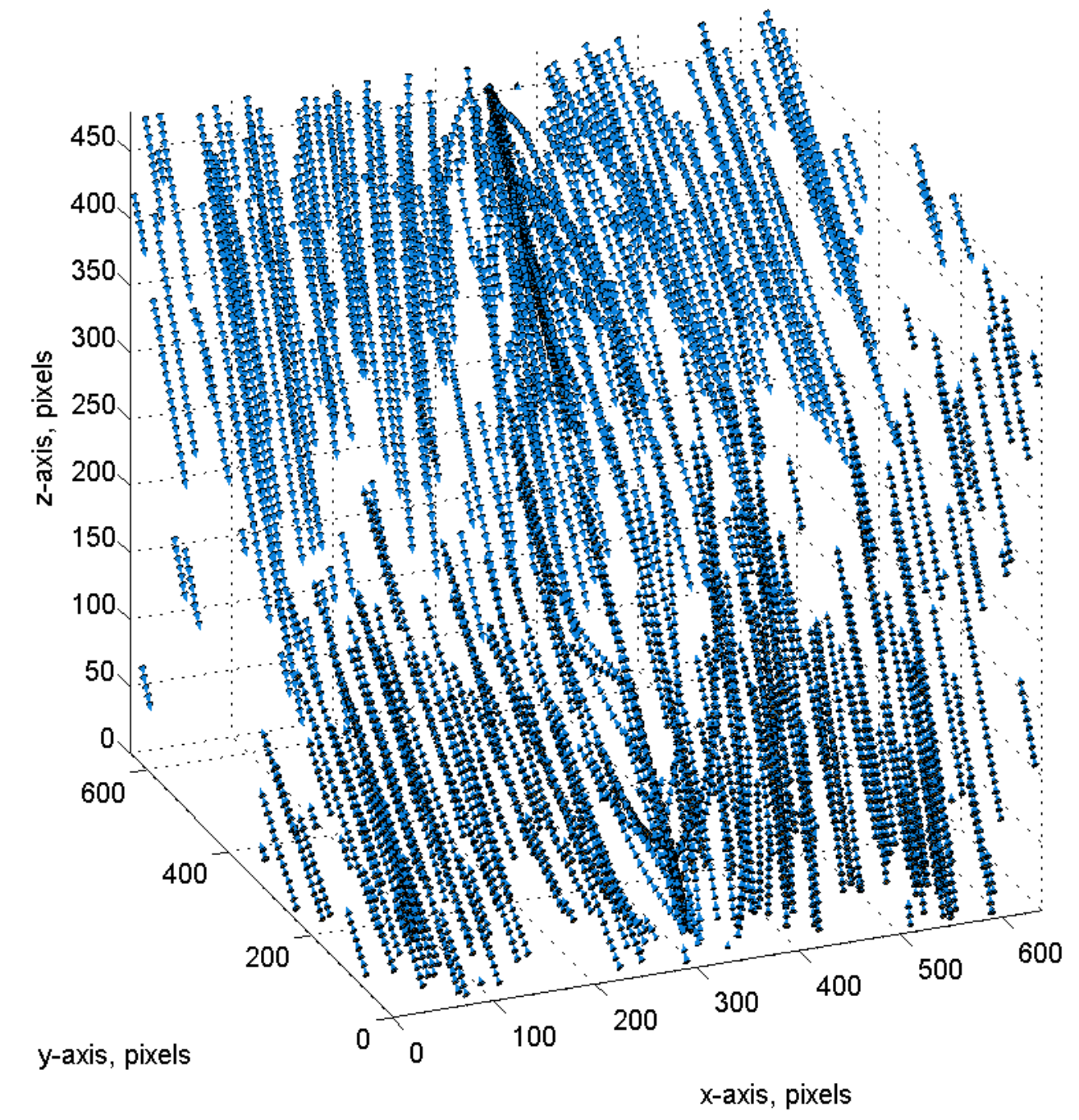}
\caption{Reconstructed velocity field obtained by tracking 500 particles in 3D over 50 frames. The arrows indicate the velocity vectors at the corresponding positions in the volume. More than 98\% of the velocity vectors are correctly reconstructed. }
\label{fig:TestII}
\end{center}
\end{figure}

The reconstructed velocity field is shown in \figref{fig:TestII}. Despite of the large number of 
ambiguities for most frames, we reconstructed more than 98$\%$ of the correct particle positions (and hence 98$\%$ of the correct velocity field). This is possible, because the objective function in~\eqref{eq:optproblem} gives preference to solutions that are consistent with our physical assumptions on the particle velocities (modeled via $\vec w^{(t)}$). 

\subsection{Camera Model}
For our simulations  we assumed a camera model for which the candidate grid was precisely determined for each frame. In Test Case~II the particle positions were simulated in real coordinates; the measurements were discretized to detector resolution. 

In real experiments, one might need to compensate for de-focusing, imperfections in the alignment, and several other effects. The projecting lines can become cones, and the opening angle may need to be estimated on an empirical basis. Large opening angles may increase the number of ghost particles. For specific ghost particle removal techniques see \cite{Novara2011, Schanz2013, Xu2008, Willneff2003, a2010}.

\section{Conclusions}\label{sect:conclusions}
Based on dynamic discrete tomography, we introduced an algorithm for 3D particle tracking velocimetry diagnostics and demonstrated its efficiency for binary projection data acquired from two projection directions. 

By using more projection directions, we can reduce the ambiguity in the solutions. This typically entails, however, that the computation time increases drastically. Exact reconstructions from two views can be particularly advantageous if the optical access is limited. Further, small PTV diagnostics are preferred due to the smaller cost. 

Although we considered only binary projection data in this paper, we remark that the algorithm can also reconstruct trajectories when more experimental data  are available (for instance, in form of brightness levels).

Another advantage of our approach is that if the sampled data intrinsically contain information that can be realized by several particle patterns, then all possible reconstruction outcomes can be generated and the correct solution can be selected \emph{post factum} based on physical principles or other information. Generation of the possible reconstruction outcomes also facilitates brief assessments of the chosen weights $\vec w$. 

\section*{Acknowledgements}
A.~A. and P.~G. were partly supported by DFG grants AL 1431/1-1, GR 993/10-1, and GR 993/10-2, respectively. D.~M. was supported by the fusion researcher fellowship program, EFDA. The authors would like to express their gratitude to M. Kantor for fruitful discussions.  COST Action MP1207 is acknowledged for networking support. The authors would like to thank the anonymous reviewers for their valuable comments and suggestions to improve the quality of the paper.

%\section*{References}
%\bibliographystyle{plain}
%\bibliography{tomobib}

\end{document}